\documentclass[fleqn,10pt]{wlscirep}
\usepackage[utf8]{inputenc}
\usepackage{setspace}
\usepackage[figuresleft]{rotating}
\usepackage[T1]{fontenc}
\title{Local and correlated studies of humidity-mediated ferroelectric thin film surface charge dynamics}
 
\author[1,2,*]{Iaroslav Gaponenko}
\author[1]{Loïc Musy}
\author[3]{Neus Domingo}
\author[4]{Nicolas Stucki}
\author[5]{Albert Verdaguer}
\author[2,6]{Nazanin Bassiri-Gharb}
\author[1]{Patrycja Paruch}
\affil[1]{Department of Quantum Matter Physics, University of Geneva, 1211, Geneva, Switzerland}
\affil[2]{G.W. Woodruff School of Mechanical Engineering, Georgia Institute of Technology, Atlanta, GA, 30332, USA}
\affil[3]{Catalan Institute of Nanoscience and Nanotechnology (ICN2), CSIC and BIST, Campus UAB, Bellaterra, 08193 Barcelona, Spain}
\affil[5]{Institut de Ci\`encia de Materials de Barcelona (ICMAB), CSIC, Campus UAB, Bellaterra, 08193 Barcelona, Spain}
\affil[4]{University of Applied Sciences and Arts Western Switzerland (HES-SO/HEPIA), 1213 Geneva, Switzerland}
\affil[6]{School of Materials Science and Engineering, Georgia Institute of Technology, Atlanta, GA, 30332, USA}

\affil[*]{iaroslav.gaponenko@unige.ch}

\begin{abstract}
Electrochemical phenomena in ferroelectrics are of particular interest for catalysis and sensing applications, with recent studies highlighting the combined role of the ferroelectric polarisation, applied surface voltage and overall switching history.
Here, we present a systematic Kelvin probe microscopy study of the effect of relative humidity and polarisation switching history on the surface charge dissipation in ferroelectric Pb(Zr$_{0.2}$Ti$_{0.8}$)O$_3$ thin films. We analyze the interaction of surface charges with ferroelectric domains through the framework of physically constrained unsupervised machine learning matrix factorization, Dictionary Learning, and reveal a complex interplay of voltage-mediated physical processes underlying the observed signal decays. Additional insight into the observed behaviours is given by a Fitzhugh-Nagumo reaction-diffusion model, highlighting the lateral spread and charge passivation process contributors within the Dictionary Learning analysis.
\end{abstract}

\begin{document}

\flushbottom
\maketitle
%
\thispagestyle{empty}

\section*{Introduction}
Surface and bulk electrochemical phenomena are particularly important in ferroelectrics, whose switchable remanent polarisation modulates their electrochemical reactivity, determining fundamental material responses and giving rise to promising catalysis and sensing applications  \cite{hwang_mattoday19_ferroelectrics_catalysis_strain, yang_natphys17_electrochemistry_nanoferro}. Electrochemistry has been linked to all aspects of ferroelectric switching and screening  \cite{neumayer_acsami_2018_lno_surface_chemistry, fabiano_sciadv17_ferro_ion_cond, ievlev_acsami18_switching_electrochemistry}, with recent studies highlighting its influence on catalytic effects and functionality-modifying phenomena \cite{kakekhani_acscata15_ferro_catalysis, garrity_prb13_ferro_surfchem}. These effects can be amplified by the additional presence of adsorbates, such as surface water ubiquitous under ambient conditions  \cite{tian_natcom18_polarization_water_printing, ievlev_acsami16_tip_induced_electrochemistry}. 

As ferroelectric surfaces exhibit varying polarisation orientation and chemical composition, the interplay of electrochemistry with surface water generates a wide range of complex phenomena. Structurally,  preferential adsorption can modulate the usual humidity dependent molecular arrangement observed on simple dielectric materials \cite{asay_jpcb05_sio2_adsorbed_water} of 1--2 monolayers of ice-like water, followed by less ordered liquid layers. Chemically, varying dissociation rates into OH$^-$ and protons vs. molecular water and even modulation of the surface and bulk material composition as a function of polarisation orientation and relative humidity have been reported  \cite{cordero_jpcc16_water_affinity_linbo3, RSCGaponenko}. 

The observed effects are further enhanced under the application of an electric field, with radical water-mediated changes and time-dependent processes taking place, as also seen in non-polar materials \cite{verdaguer_apl_2009_graphene_charging_kpfm}. In addition, sufficiently high fields will enhance oxidation reactions at the surface and promote charge injection and/or polarisation switching \cite{segura_jap_2013_domain_surface_screening, ievlev_apl14_humidity_linbo3, ievlev_natphys13_chaotic_switching}, leading to a strong dependence of electrochemical response on the electrostatic (switching and charging) history of the material. The combination of all these phenomena gives rise to a highly nontrivial mixed response of ferroelectrics to changes in relative humidity or surface bias, with widely ranging timescales and magnitudes, as demonstrated previously by local probe surface potential imaging \cite{RSCGaponenko, shishkin_apl06_lead_germanate_pfm}.

Over the last decade, investigation of complex phenomena with multiple parameters have been made possible by incremental advances in experimental techniques and acquisition hardware, yielding rich datasets of correlated responses at a broad range of timescales. Additionally, tremendous progress has been made in tools designed to analyze the resulting hyperdimensional datasets, based on Big Data approaches such as machine learning and dimensional reduction. These have been successfully applied to studies of ferroelectric materials, enabling a better understanding of the competing and correlated processes involved in switching and surface screening under ambient conditions \cite{vasudevan_jap15_pca_relaxor_relaxation, li_sciadv18_machine_learning_phase_transitions, griffin_npjcompmat19_smart_machine_learning}.

Here, we investigate the mechanisms of surface charge dissipation on ferroelectric thin film surfaces as a function of time, applied voltage, writing history from as-grown through final polarisation, and relative humidity. Local probe imaging of the surface potential was performed on two Pb(Zr$_{0.2}$Ti$_{0.8}$)O$_3$ (PZT) thin films of opposite polarisation, after the writing of a predefined structure. The resulting time-dependent datasets were dimensionally-stacked and analyzed by means of Dictionary Learning in order to unravel the underlying physical and chemical behaviours. A simplified reaction-diffusion model complements the experimental work, allowing the results of the machine learning analysis to be interpreted in terms of physical/chemical processes. Our observations unequivocally demonstrate that whilst both positive and negative surface charges decay with time, only negative charges spread laterally on the surface, diffusing far beyond the originally written areas.

\section*{Results and discussion}
\subsection*{Imaging surface charge dissipation}
\begin{figure}[ht]
\centering
\includegraphics[width=0.75\linewidth]{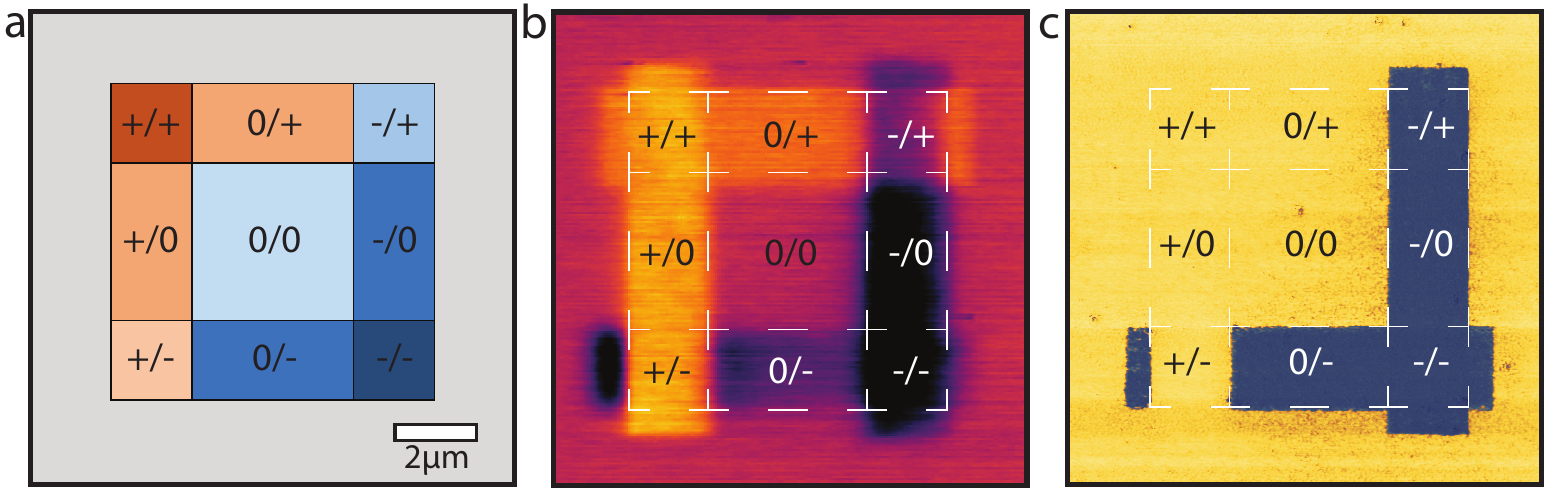}
\caption{Experimental protocol for mapping charge dissipation on the ferroelectric thin film surface. In each sample and at each humidity setpoint, a biased atomic force microscope tip was scanned over the surface: first defining lateral stripe regions with positive (+8 V), zero, and negative bias (-8 V) at 0$^\circ$; then defining vertical stripe regions with identical positive, zero, and negative bias at 90$^\circ$. (a) Schema of the target writing pattern. (b) Typical surface potential (KPFM) image of the resulting structure, with overlaid writing bias pattern. (c) Ferroelectric domain structure resulting from voltage application on the down-polarised sample (PFM phase image).}
\label{fig:protocol}
\end{figure}

Surface charge dissipation was imaged by Kelvin probe force microscopy (KPFM) on 100 nm thick PZT films (presenting either up- or down-oriented as-grown monodomain polarisation) and tracked as a function of relative humidity, polarisation, applied voltages, and time. Domain structures were patterned by scanning the biased atomic force microscope tip in contact with the surface: first, to define three lateral stripe regions with positive (+8 V), zero, and negative (-8 V) tip bias; then, three overlaid vertical stripe regions, again with positive, zero, and negative tip bias, as illustrated in Fig. \ref{fig:protocol} (a). The sequential writing is performed in order to take into account possible effects of polarisation switching history and varying intensity of charge injection (areas scanned twice with negative or positive voltages). Once the writing was completed, KPFM imaging was carried out until dissipation of the surface potential contrast, approximately 9 to 12 hours. An illustration of resulting KPFM image is shown in Fig. \ref{fig:protocol} (b), with the writing structure overlaid. We note that whilst the piezoresponse force microscopy image of the ferroelectric domains generated through the writing process follows the writing structure exactly, the KPFM usually extends outwards due to charge diffusion, especially at high humidity. \footnote{We note that a 20\% overscan generated by the atomic force microscope used in this study, required for the operation of the closed loop scanner, results in the extension of the stripe regions along their length, giving rise to additional sharp contrast regions outside the formally defined writing structure.} The procedure was repeated at each relative humidity setpoint (ten different values spanning 6\% to 80\%), for both samples (up- and down-oriented as-grown polarisation). The control of relative humidity was performed with a in-house-built flow-based, low noise humidity controller, which enables fast and precise \textit{in-situ} control in the atomic force microscope chamber \cite{gaponenko_rsi16_humidity_controller, gaponenko_erx19_humidity_controller}.

\begin{figure}[ht]
\centering
\includegraphics[width=0.5\linewidth]{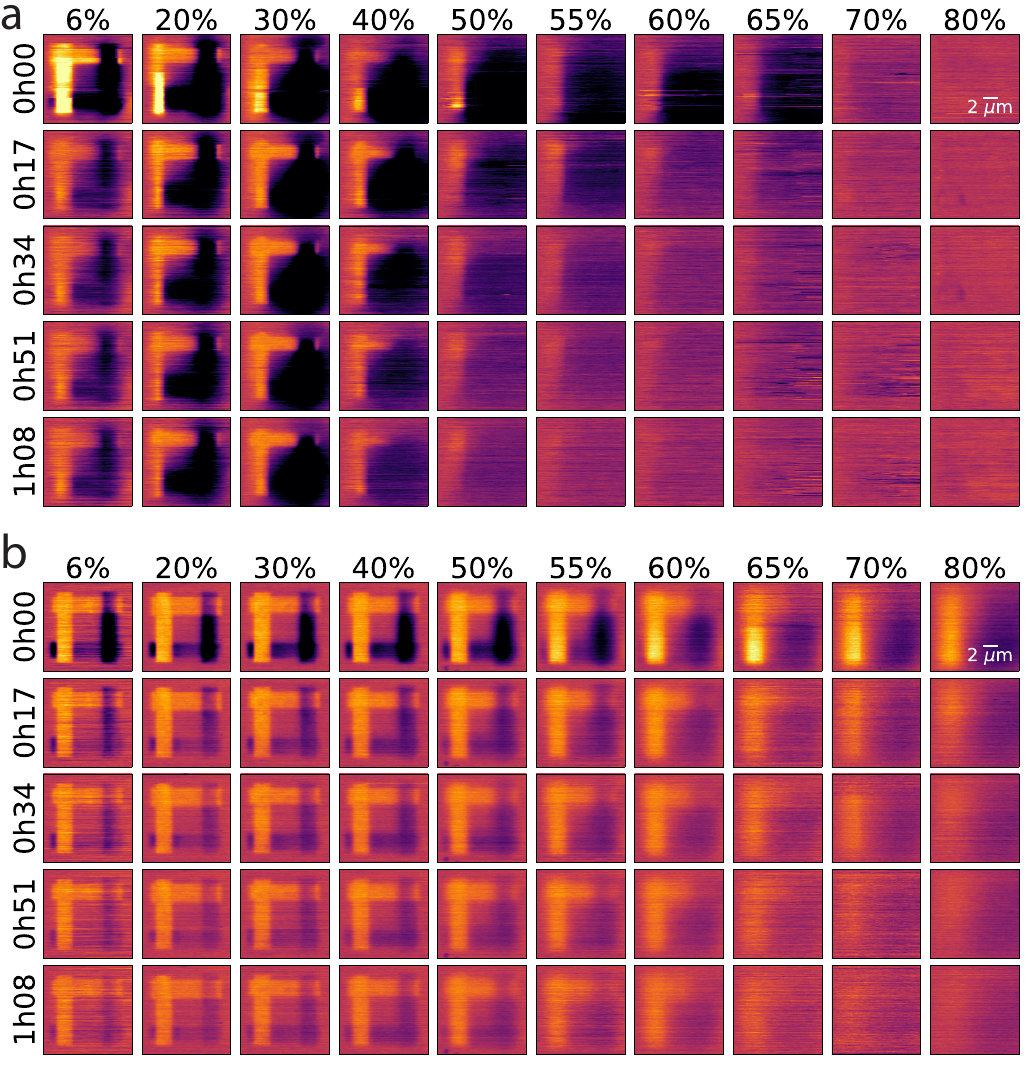}
\caption{First five KPFM images for the complete humidity dataset, as a function of time and humidity for the (a) down-polarised and (b) up-polarised samples. A new structure is written at each relative humidity setpoint and then monitored continuously by KPFM until complete surface potential dissipation. At low humidities the charges seem to be more localized and retained longer than at high humidities. Additionally, at higher humidities the horizontal lines corresponding to the first writing step are not visible, suggesting much faster charge dynamics in presence of high water content at the surface. Lastly, the written pattern remains visible in the up-polarised films at much higher humidity levels than in the down-polarised films.}
\label{fig:kpfm}
\end{figure}

The first five resulting surface potential images (approximately 70 minutes after writing) are shown in Fig. \ref{fig:kpfm} for both samples at all ten humidity levels probed. Several trends can be identified from simple visual inspection. First, in both samples, faster surface charge dissipation and larger spatial spread of the KPFM signal beyond the boundaries of the stripe domains can be observed at higher humidities. Second, the positive charges seem to be much more localized (bright yellow contrast) than the negative charges (dark blue contrast), regardless of the initial sample polarisation. Finally, the charge contrast seems to be more intense in areas where the polarisation was switched from its as-grown state - i.e. where negative tip bias was applied for the down-polarised film and positive tip bias for the up-polarised film - hinting at the role of switching history in the dissipation. These observations are in line with previous investigations of the dynamics of charged species on ferroelectric thin films \cite{RSCGaponenko}, showing a strong effect of relative humidity on charge dissipation. However, the additional parameter of history yields new insight: in both samples the first writing (horizontal stripes) can be readily distinguished at low humidities, but is not discernible at high humidities, suggesting faster dynamics in presence of high water content at the surface.

\subsection*{Dictionary Learning analysis}
Due to the complex and correlated nature of the full dataset, any analysis beyond simple visual inspection is challenging as it will introduce human error and bias: regions over which observed behaviour is to be averaged may be selected arbitrarily, outliers and/or unexpected data might be ignored, and overall the spatial distribution of information within the images will be lost as most analysis would concentrate on average signals. To address these issues, machine learning-based techniques were leveraged for the correlative analysis of the acquired data.

The KPFM images for all ten humidities and two samples (one up-polarised and one down-polarised) were stacked along the spatial dimension in order to induce a time correlation, and are shown in the Supplementary Materials, Figure S1. Such stacking imposes a constraint of identical behaviour(s) across polarisation, time and humidity values \cite{griffin_npjcompmat19_smart_machine_learning}, thus enabling a direct comparison of the evolution of the behaviours as a function of initial sample polarisation and relative humidity. Dictionary Learning was selected as the machine learning dimensional reduction technique for the further analysis of the dataset due to its intrinsically sparse nature \cite{Dumitrescu2018_dict_learning}, with the expectation of only a few behaviours being spatially co-located. The number of components $n$ is a free parameter and was evaluated from two to ten, allowing us to track which behaviours are associated with the different regions as they successively segregate out, as detailed in the Supplementary Materials, Figure S3. 

\begin{figure}[ht]
\centering
\includegraphics[width=\linewidth]{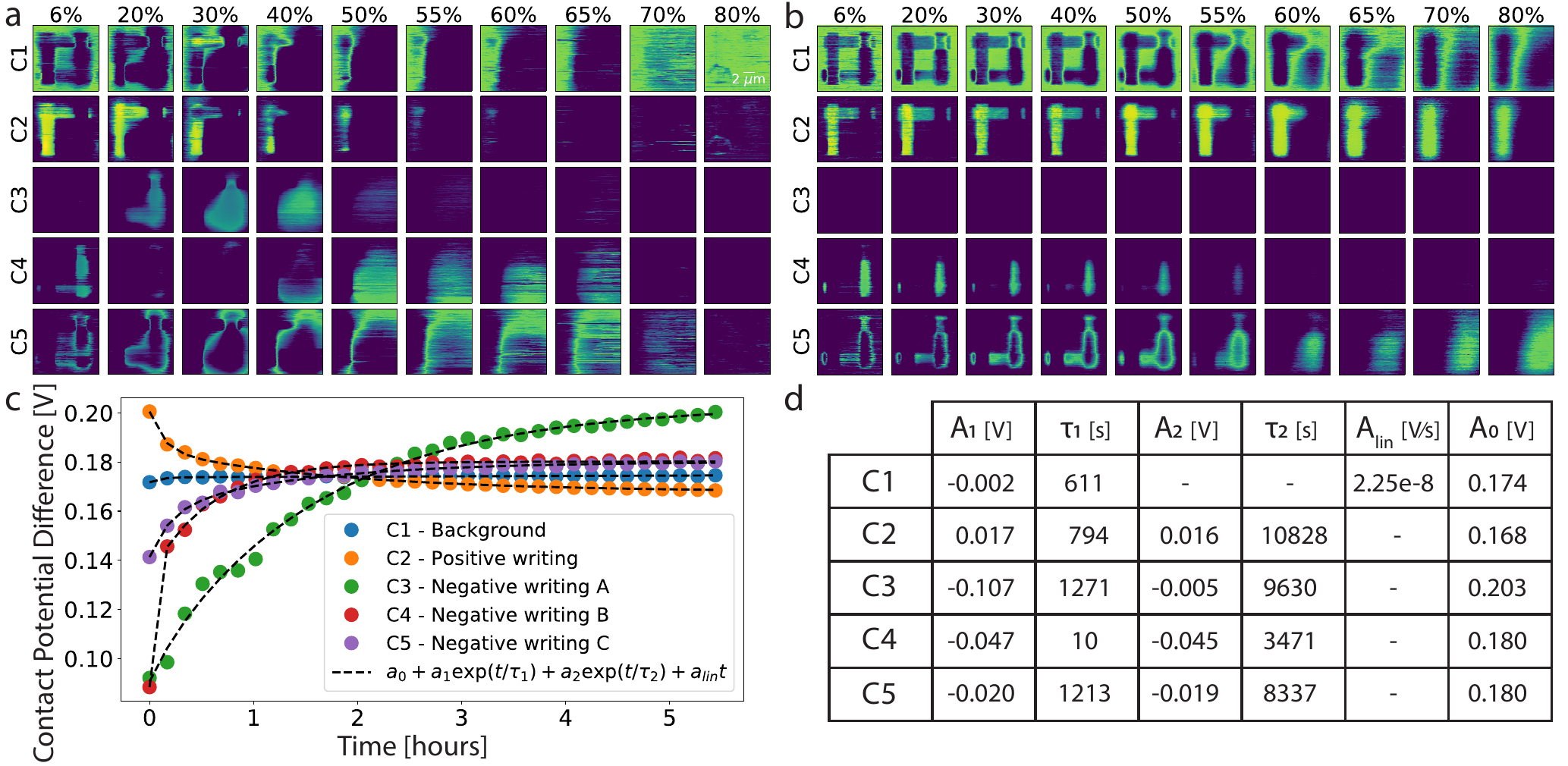}
\caption{Dictionary Learning spectral decomposition of the complete dataset of surface potential evolution measured in the two samples and at ten humidities into five components. The weight maps for the (a) down-polarised and (b) up-polarised sample demonstrate spatially localized features, which once correlated with the (c) time dependent components yield a clear separation between the background response (\textit{C1}), and the areas of positive (\textit{C2}) and negative (\textit{C3-5}) charge. The negative charge region is split into co-located concentric structures, reminiscent of Matryoshka dolls. Components C2 -- C5 can be fitted with double exponential decays, each yielding (d) two very different time constants.}

\label{fig:dlanalysis}
\end{figure}

The resulting decomposition, here presented for $n = 10$, shows three distinct types of components, each associated with the relevant spatial distribution weight maps. The first component, C1, has a low magnitude and is spatially located in the regions that can be identified as a background - i.e., locations outside of the written structure or where zero tip bias was applied during writing - and is identifiable as a distinct component for $n\geq8$. This component is also present within the written areas at higher humidities, and specifically in the locations where the surface charge has already dissipated. An inspection of its time-evolution indicates that this component is essentially stable with a very slight linear drift, possibly caused by the change in tip-sample interactions over time, as shown in the exponential-plus-linear fit C1 in Fig. \ref{fig:dlanalysis}d. The second component, C2, decays with time from an initial high positive value, and is present in areas of positive voltage application only, regardless of the initial polarisation state of the films. These areas separate out as distinct from those written with negative bias at $n=6$. Increasing $n$ does not change the C2 component. The last three components present negative decays, starting at a low surface potential value and evolving towards the background potential as the surface charges dissipate. These three components are co-located within each other and spread outwards around areas of negative voltage application, with the highest intensity decay (C3) on the inside, and the lowest intensity decay (C5) on the outside of the negative charge distributions - a structure reminiscent of Matryoshka dolls, and suggesting the possible presence of lateral diffusion. This characteristic structure is apparent from $n=7$, and increasing $n$ simply refines the areas written with negative bias into increasingly fine concentric segments. 

To quantify the time evolution of the positive and negative decays, the corresponding components - C2 to C5 - were fitted to a double exponential decay function, $CPD(t) = A_0+A_1\exp{t/\tau_1}+A_2\exp{t/\tau_2}$, with $CPD$ the contact potential difference, and $A_0$ the offset corresponding to the final CPD value, $A_1, A_2$ the amplitudes, and $\tau_1, \tau_2$ the time constant of the decays. From the parameters shown in Fig. \ref{fig:dlanalysis}d, two behaviors seem to be present in all decaying components, each with two substantially different time constants. The latter differ by at least an order of magnitude, as reported previously for this material, indicating the coexistence of a material-dependent fast electrochemical process and a slow process mediating the surface charge dissipation \cite{RSCGaponenko}. The fast process, of the order of 1000 seconds, can be related to the re-equilibration of surface screening charges \cite{Tong2015}, whereas the slow process, of the order of 10000 seconds, is specifically attributed to the dynamics of tip-injected surface or near-surface charges.

In all cases, however, the overall mechanisms (background, positive and negative decay) are present in both films and share the same parameters, varying in intensity with humidity and initial polarisation, but mostly depending on the polarity of the applied voltage. The positive voltage related component, C2, for instance is found in both films below 50\% RH, but seems to be more present at the higher humidity only in the up-polarised films, where a positive voltage application switches the polarisation. Conversely, negative voltage related components, C3, C4, and C5 seem to be more present within the down-polarised film, where such a voltage polarity switches the polarisation. The coexistence of the same physical dissipation processes (as fitted above) in both samples suggests that the initial polarisation orientation does not affect the dissipation mechanisms which seem to be primarily dominated by voltage and surface electrochemistry. However, the \textit{history} of polarisation is important, as switching processes during voltage application will result in higher surface charge density and/or longer charge retention, especially at higher humidities.
  
\subsection*{Reaction-diffusion simulation}
To better understand the specific behaviors observed in the experimental data and the resulting Machine Learning analysis, a simplified reaction-diffusion model was constructed, based on the FitzHugh-Nagumo system. This is an excitable system model, ideal for describing the distributions of different species during their relaxation to equilibrium, as for example in neural synapses \cite{Rocoreanu2000_fitzhugh_nagumo}. As the experimental data suggests the presence of lateral spread acting together with surface and bulk electrochemistry, the parameters of the model were tuned to reproduce the observations qualitatively, with a spatially localized and negligibly diffusive positive charge distribution, and an initially more spread and highly diffusive negative charge distribution.

\begin{figure}[ht]
\centering
\includegraphics[width=\linewidth]{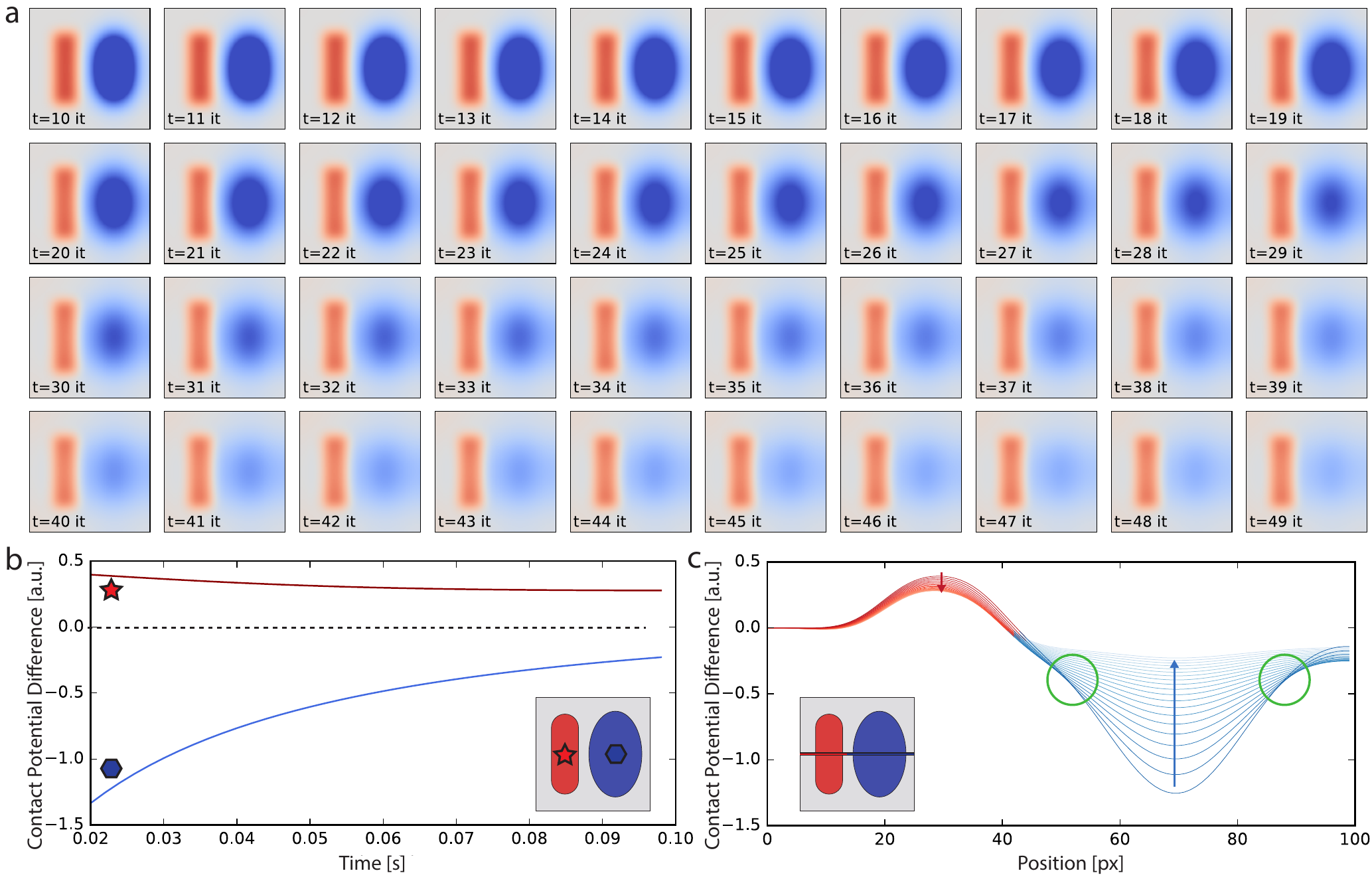}
\caption{FitzHugh-Nagumo reaction-diffusion model of the surface charge dissipation. (a) Time evolution of the positive (red) and negative (blue) surface charge distributions shows spatially localised vs. highly diffusive behavior, qualitatively reproducing the experimental observations. Maxima (b) and cross-sections (c) of the positive and negative charge distributions show that both positive and negative charge intensities decay with time, but that only negative charges show a lateral diffusion, as evidenced by the increased width of the signal, leading to characteristic crossing points (green circles).}
\label{fig:rdresults}
\end{figure}

The resulting time evolution of the combined distributions is shown in Fig. \ref{fig:rdresults}a. As a function of time, the intensity of both distributions appears to decay, as illustrated in the plot of the evolution of each distribution's center point intensity. However, the difference in diffusivity of the two distributions is apparent when a cross-section is taken at the center of each image. Whilst the positive charges (red) seem to simply decay (i.e. reduce in intensity as a function of time), the negative charge distribution spreads outwards via lateral diffusion, as indicated by the characteristic crossing points.

\begin{figure}[ht]
\centering
\includegraphics[width=0.5\linewidth]{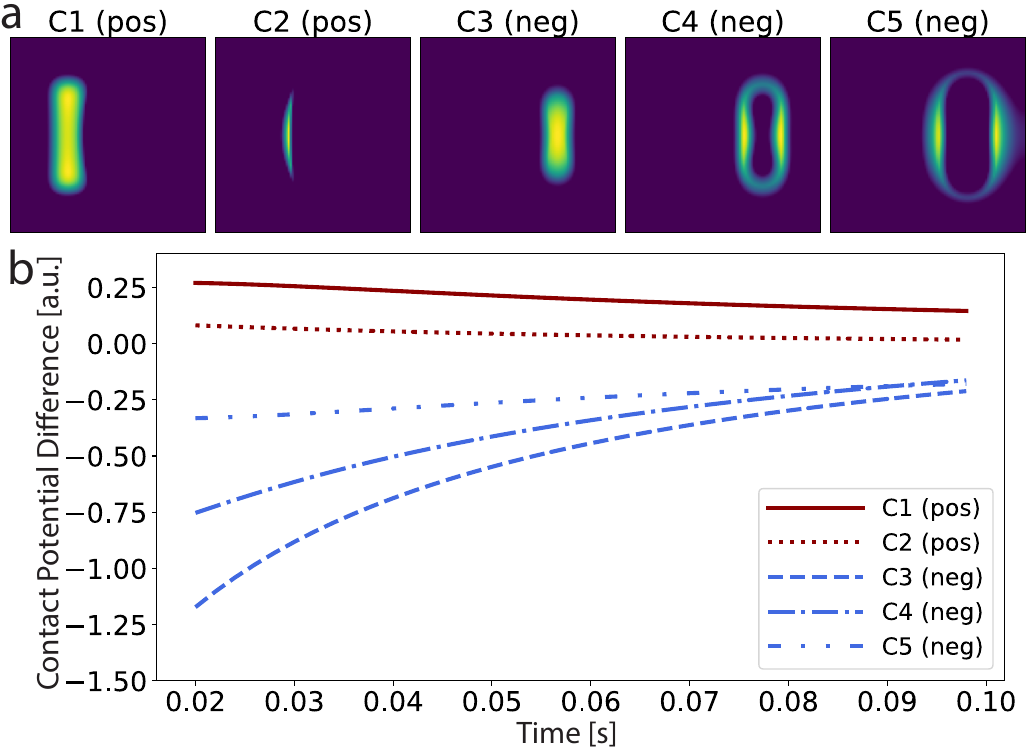}
\caption{Dictionary Learning analysis of Reaction-Diffusion analysis. The maps shown in (a) and corresponding components shown in (b) demonstrate two specific behaviours: positive and negative decays. The latter show a decreasing amplitude following a characteristic concentric structure, consistent with the presence of a diffusion process.} 
\label{fig:rddictlearn}
\end{figure}

The model data were subjected to the same computational treatment as the experimental data with Dictionary Learning analysis. The resulting components and weight maps are shown in Fig. \ref{fig:rddictlearn}, separated into behaviors for positively and negatively charged regions. For the positive regions, only simple decays are observed, with the second component identifying mostly the interaction of charges at the boundary between positively and negatively charged regions. In comparison, the negative charge decay shows a striking resemblance to the experimental data analysis, with concentric shell structures in the weight maps.

Lastly, the components associated with the weight maps are also consistent with the original experimental observations: the most and least intense behaviours are located at the center and at the periphery of each charged regions, respectively. Per model parameters, the species that shows this behavior is associated with a reaction and diffusion process - as opposed to the other species which is constrained to reaction only. Thus, the introduction of diffusion into the model brings out the characteristic concentric shell structure of decreasing intensity in the Dictionary Learning analysis. This result confirms the intuition that lateral diffusion is taking place within our experimental data based on the similarity to the machine learning analysis features, making this - to the authors' knowledge - the first reported observation of a signature of such a behaviour observed by machine learning methods.

\section*{Conclusion}
In conclusion, in this investigation of the interplay of time, relative humidity, polarisation and voltage with the surface charge dissipation of ferroelectric thin films, polarised regions written by scanning probe microscopy were analyzed with Dictionary Learning spectral decomposition. The results highlight two separate voltage-mediated behaviours, characteristic of a combined surface/bulk electrochemical process and of surface charge passivation. Negative charges are additionally observed to diffuse laterally, as demonstrated by the characteristic concentric shell structure in machine learning analysis, while positive charges are mostly confined to their original location with very limited lateral diffusion. Our interpretation is corroborated by a reaction-diffusion model set up to reproduce the experimental KPFM observations, whose resulting time evolution was analyzed with the same spectral decomposition technique, yielding similar characteristic concentric structures in the case of the diffusing species.
The cumulative evidence points to a non-trivial behaviour of surface charges, with the positive and negative charges exhibiting a substantially different diffusion/dissipation mechanism that is modulated by the underlying ferroelectric polarisation. The coexistence of these electrochemical processes opens a pathway towards polarisation-directed catalysis, with a highly localized control of charges.

\section*{Materials and methods}
\subsection*{Materials growth}
Films of 100 nm thick Pb(Zr$_{0.2}$Ti$_{0.8}$)O$_3$ (PZT) have been grown on Nb:SrTiO$_3$ single crystalline substrates by RF-magnetron sputtering, as described elsewhere \cite{gariglio_apl_2007_pzt_growth}. The films used in this study were up- and down-polarised as grown, respectively, with a smooth sub-nanometer roughness topography.

\subsection*{Kelvin Probe Force Microscopy}
The scanning probe microscopy imaging was performed on an \textit{Asylum Research Cypher} microscope equipped with an environmental control chamber, using \textit{Bruker SCM-PtSi} cantilevers with a resonance frequency of $\sim 75kHz$ and a spring constant of $\sim 2.8 N/m$. Sample temperature was kept constant at $30^\circ C$, and relative humidity was varied through the use of a home-build controller \cite{gaponenko_rsi16_humidity_controller, gaponenko_erx19_humidity_controller}. The humidity was stabilized for at least an hour prior to each measurement series. Topographic imaging was performed in intermittent contact mode (\textit{AC Mode}), with a free oscillation amplitude of $\sim 10 nm$ and a setpoint of $\sim 5 nm$. The contact potential difference between the tip and the sample was acquired over $12x12\mu m$ areas at a scan speed of $0.5Hz$ in single pass Kelvin probe force microscopy mode with the following parameters: $3V$ excitation voltage, $5kHz$ frequency, $-90^\circ$ phase, $5$ proportional gain, $10$ integral gain. Domain writing was done in contact mode with a $0.5V$ setpoint over $8x8\mu m$ areas at a scan speed of $0.5Hz$ and applied voltages of $\pm 8V$, as detailed in Fig. \ref{fig:protocol}(a).

\subsection*{Dictionary Learning analysis}
Time-dependent data was stacked into data cubes, and the data cubes at each humidity and for each sample were concatenated along the spatial dimension \textit{x}, yielding a correlated dataset. A similar stacking method was performed for the reaction-diffusion modeling dataset. The Dictionary Learning implementation of the \textit{scikit-learn} library was then used on the flattened data set with the following parameters: $3-10$ n\_components, $1000000$ max\_iter, $8$ n\_jobs, \textit{lasso\_lars} transform\_algorithm, $15$ transform\_n\_nonzero\_coefs, \textit{lars} fit\_algorithm, and $0$ random\_state. The code and data used for the analysis is freely available on the paper repository \textbf{[CITE yareta\_or\_zenodo]}.

\subsection*{Reaction-diffusion modeling}
The reaction-diffusion model used is based on a 2D Fitzhugh-Nagumo excitable system model \cite{Rocoreanu2000_fitzhugh_nagumo}. The implementation used is adapted from existing Python code, recipe 12.4 in \cite{rossant2018ipython}. The specific form of the model is a set of two differential equations:
$$\frac{\partial u}{\partial t} = a\Delta u + u - u^3 - v + k$$
$$\tau \frac{\partial v}{\partial t} = b\Delta v + u - v$$
with $u,v$ denoting the two species, $a,b$ their corresponding diffusion factors, $k$ the intensity of the external stimulus, and $tau$ the time constant providing the separation of the timescales of the evolution of the two species. For the comparison to surface charge dissipation, the two species were initially set up in separate bands and allowed to relax. The parameter set, selected up to allow for single species diffusion only, is: $a=0$, $b=0.1$, $\tau = 0.1$, and $k= -.005$. The resulting spatiotemporal evolution was analyzed through the same Dictionary Learning workflow as the experimetnally obtained data set. The code and data used for the simulation and analysis is freely available on the paper repository \textbf{[CITE yareta\_or\_zenodo]}.

\bibliography{sample}

\begin{thebibliography}{10}
\urlstyle{rm}
\expandafter\ifx\csname url\endcsname\relax
  \def\url#1{\texttt{#1}}\fi
\expandafter\ifx\csname urlprefix\endcsname\relax\def\urlprefix{URL }\fi
\expandafter\ifx\csname doiprefix\endcsname\relax\def\doiprefix{DOI: }\fi
\providecommand{\bibinfo}[2]{#2}
\providecommand{\eprint}[2][]{\url{#2}}

\bibitem{hwang_mattoday19_ferroelectrics_catalysis_strain}
\bibinfo{author}{Hwang, J.} \emph{et~al.}
\newblock \bibinfo{journal}{\bibinfo{title}{Tuning perovskite oxides by strain:
  Electronic structure, properties, and functions in (electro)catalysis and
  ferroelectricity}}.
\newblock {\emph{\JournalTitle{Materials Today}}}
  \doiprefix\url{https://doi.org/10.1016/j.mattod.2019.03.014}
  (\bibinfo{year}{2019}).

\bibitem{yang_natphys17_electrochemistry_nanoferro}
\bibinfo{author}{Yang, S.~M.} \emph{et~al.}
\newblock \bibinfo{journal}{\bibinfo{title}{Mixed
  electrochemical{\textendash}ferroelectric states in nanoscale
  ferroelectrics}}.
\newblock {\emph{\JournalTitle{Nature Physics}}} \textbf{\bibinfo{volume}{13}},
  \bibinfo{pages}{812--818}, \doiprefix\url{10.1038/nphys4103}
  (\bibinfo{year}{2017}).

\bibitem{neumayer_acsami_2018_lno_surface_chemistry}
\bibinfo{author}{Neumayer, S.~M.} \emph{et~al.}
\newblock \bibinfo{journal}{\bibinfo{title}{{Surface Chemistry Controls
  Anomalous Ferroelectric Behavior in Lithium Niobate}}}.
\newblock {\emph{\JournalTitle{ACS Applied Materials \& Interfaces}}}
  \textbf{\bibinfo{volume}{{10}}}, \bibinfo{pages}{{29153--29160}},
  \doiprefix\url{{10.1021/acsami.8b09513}} (\bibinfo{year}{{2018}}).

\bibitem{fabiano_sciadv17_ferro_ion_cond}
\bibinfo{author}{Fabiano, S.} \emph{et~al.}
\newblock \bibinfo{journal}{\bibinfo{title}{Ferroelectric polarization induces
  electronic nonlinearity in ion-doped conducting polymers}}.
\newblock {\emph{\JournalTitle{Science Advances}}}
  \textbf{\bibinfo{volume}{3}}, \doiprefix\url{10.1126/sciadv.1700345}
  (\bibinfo{year}{2017}).
\newblock
  \eprint{https://advances.sciencemag.org/content/3/6/e1700345.full.pdf}.

\bibitem{ievlev_acsami18_switching_electrochemistry}
\bibinfo{author}{Ievlev, A.~V.} \emph{et~al.}
\newblock \bibinfo{journal}{\bibinfo{title}{Nanoscale electrochemical phenomena
  of polarization switching in ferroelectrics}}.
\newblock {\emph{\JournalTitle{ACS Applied Materials \& Interfaces}}}
  \textbf{\bibinfo{volume}{10}}, \bibinfo{pages}{38217--38222},
  \doiprefix\url{10.1021/acsami.8b13034} (\bibinfo{year}{2018}).
\newblock \eprint{https://doi.org/10.1021/acsami.8b13034}.

\bibitem{kakekhani_acscata15_ferro_catalysis}
\bibinfo{author}{Kakekhani, A.} \& \bibinfo{author}{Ismail-Beigi, S.}
\newblock \bibinfo{journal}{\bibinfo{title}{Ferroelectric-based catalysis:
  Switchable surface chemistry}}.
\newblock {\emph{\JournalTitle{{ACS} Catalysis}}} \textbf{\bibinfo{volume}{5}},
  \bibinfo{pages}{4537--4545}, \doiprefix\url{10.1021/acscatal.5b00507}
  (\bibinfo{year}{2015}).

\bibitem{garrity_prb13_ferro_surfchem}
\bibinfo{author}{Garrity, K.}, \bibinfo{author}{Kakekhani, A.},
  \bibinfo{author}{Kolpak, A.} \& \bibinfo{author}{Ismail-Beigi, S.}
\newblock \bibinfo{journal}{\bibinfo{title}{Ferroelectric surface chemistry:
  First-principles study of the pbtio${}_{3}$ surface}}.
\newblock {\emph{\JournalTitle{Phys. Rev. B}}} \textbf{\bibinfo{volume}{88}},
  \bibinfo{pages}{045401}, \doiprefix\url{10.1103/PhysRevB.88.045401}
  (\bibinfo{year}{2013}).

\bibitem{tian_natcom18_polarization_water_printing}
\bibinfo{author}{Tian, Y.} \emph{et~al.}
\newblock \bibinfo{journal}{\bibinfo{title}{Water printing of ferroelectric
  polarization}}.
\newblock {\emph{\JournalTitle{Nature Communications}}}
  \textbf{\bibinfo{volume}{9}}, \doiprefix\url{10.1038/s41467-018-06369-w}
  (\bibinfo{year}{2018}).

\bibitem{ievlev_acsami16_tip_induced_electrochemistry}
\bibinfo{author}{Ievlev, A.~V.} \emph{et~al.}
\newblock \bibinfo{journal}{\bibinfo{title}{Chemical state evolution in
  ferroelectric films during tip-induced polarization and electroresistive
  switching}}.
\newblock {\emph{\JournalTitle{ACS Applied Materials \& Interfaces}}}
  \textbf{\bibinfo{volume}{8}}, \bibinfo{pages}{29588--29593},
  \doiprefix\url{10.1021/acsami.6b10784} (\bibinfo{year}{2016}).
\newblock \eprint{https://doi.org/10.1021/acsami.6b10784}.

\bibitem{asay_jpcb05_sio2_adsorbed_water}
\bibinfo{author}{Asay, D.~B.} \& \bibinfo{author}{Kim, S.~H.}
\newblock \bibinfo{journal}{\bibinfo{title}{Evolution of the adsorbed water
  layer structure on silicon oxide at room temperature}}.
\newblock {\emph{\JournalTitle{The Journal of Physical Chemistry B}}}
  \textbf{\bibinfo{volume}{109}}, \bibinfo{pages}{16760--16763},
  \doiprefix\url{10.1021/jp053042o} (\bibinfo{year}{2005}).
\newblock \bibinfo{note}{PMID: 16853134},
  \eprint{https://doi.org/10.1021/jp053042o}.

\bibitem{cordero_jpcc16_water_affinity_linbo3}
\bibinfo{author}{Cordero-Edwards, K.} \emph{et~al.}
\newblock \bibinfo{journal}{\bibinfo{title}{Water affinity and surface charging
  at the z-cut and y-cut linbo3 surfaces: An ambient pressure x-ray
  photoelectron spectroscopy study}}.
\newblock {\emph{\JournalTitle{The Journal of Physical Chemistry C}}}
  \textbf{\bibinfo{volume}{120}}, \bibinfo{pages}{24048--24055},
  \doiprefix\url{10.1021/acs.jpcc.6b05465} (\bibinfo{year}{2016}).
\newblock \eprint{https://doi.org/10.1021/acs.jpcc.6b05465}.

\bibitem{RSCGaponenko}
\bibinfo{author}{domingo, n.} \emph{et~al.}
\newblock \bibinfo{journal}{\bibinfo{title}{Surface charged species and
  electrochemistry of ferroelectric thin films}}.
\newblock {\emph{\JournalTitle{Nanoscale}}} \bibinfo{pages}{--},
  \doiprefix\url{10.1039/C9NR05526F} (\bibinfo{year}{2019}).

\bibitem{verdaguer_apl_2009_graphene_charging_kpfm}
\bibinfo{author}{Verdaguer, A.} \emph{et~al.}
\newblock \bibinfo{journal}{\bibinfo{title}{{Charging and discharging of
  graphene in ambient conditions studied with scanning probe microscopy}}}.
\newblock {\emph{\JournalTitle{Applied Physics Letters}}}
  \textbf{\bibinfo{volume}{{94}}}, \doiprefix\url{{10.1063/1.3149770}}
  (\bibinfo{year}{{2009}}).

\bibitem{segura_jap_2013_domain_surface_screening}
\bibinfo{author}{Segura, J.~J.}, \bibinfo{author}{Domingo, N.},
  \bibinfo{author}{Fraxedas, J.} \& \bibinfo{author}{Verdaguer, A.}
\newblock \bibinfo{journal}{\bibinfo{title}{{Surface screening of written
  ferroelectric domains in ambient conditions}}}.
\newblock {\emph{\JournalTitle{Joural of Applied Physics}}}
  \textbf{\bibinfo{volume}{{113}}}, \doiprefix\url{{10.1063/1.4801983}}
  (\bibinfo{year}{{2013}}).
\newblock \bibinfo{note}{{21st IEEE International Symposium on Applications of
  Ferroelectrics held jointly with 11th European Conference on the Applications
  of Polar Dielectrics and 4th Conference on Piezoresponse Force Microscopy and
  Nanoscale Phenomena in Polar Materials, Univ Aveiro, Aveiro, PORTUGAL, JUL
  09-13, 2012}}.

\bibitem{ievlev_apl14_humidity_linbo3}
\bibinfo{author}{Ievlev, A.~V.}, \bibinfo{author}{Morozovska, A.~N.},
  \bibinfo{author}{Shur, V.~Y.} \& \bibinfo{author}{Kalinin, S.~V.}
\newblock \bibinfo{journal}{\bibinfo{title}{Humidity effects on tip-induced
  polarization switching in lithium niobate}}.
\newblock {\emph{\JournalTitle{Applied Physics Letters}}}
  \textbf{\bibinfo{volume}{104}}, \bibinfo{pages}{092908},
  \doiprefix\url{10.1063/1.4867979} (\bibinfo{year}{2014}).
\newblock \eprint{https://doi.org/10.1063/1.4867979}.

\bibitem{ievlev_natphys13_chaotic_switching}
\bibinfo{author}{Ievlev, A.~V.} \emph{et~al.}
\newblock \bibinfo{journal}{\bibinfo{title}{Intermittency, quasiperiodicity and
  chaos in probe-induced ferroelectric domain switching}}.
\newblock {\emph{\JournalTitle{Nature Physics}}} \textbf{\bibinfo{volume}{10}},
  \bibinfo{pages}{59--66}, \doiprefix\url{10.1038/nphys2796}
  (\bibinfo{year}{2013}).

\bibitem{shishkin_apl06_lead_germanate_pfm}
\bibinfo{author}{Shishkin, E.~I.}, \bibinfo{author}{Shur, V.~Y.},
  \bibinfo{author}{Schlaphof, F.} \& \bibinfo{author}{Eng, L.~M.}
\newblock \bibinfo{journal}{\bibinfo{title}{Observation and manipulation of the
  as-grown maze domain structure in lead germanate by scanning force
  microscopy}}.
\newblock {\emph{\JournalTitle{Applied Physics Letters}}}
  \textbf{\bibinfo{volume}{88}}, \bibinfo{pages}{252902},
  \doiprefix\url{10.1063/1.2183369} (\bibinfo{year}{2006}).
\newblock \eprint{https://doi.org/10.1063/1.2183369}.

\bibitem{vasudevan_jap15_pca_relaxor_relaxation}
\bibinfo{author}{Vasudevan, R.~K.} \emph{et~al.}
\newblock \bibinfo{journal}{\bibinfo{title}{Multidimensional dynamic
  piezoresponse measurements: Unraveling local relaxation behavior in
  relaxor-ferroelectrics via big data}}.
\newblock {\emph{\JournalTitle{Journal of Applied Physics}}}
  \textbf{\bibinfo{volume}{118}}, \bibinfo{pages}{072003},
  \doiprefix\url{10.1063/1.4927803} (\bibinfo{year}{2015}).
\newblock \eprint{https://doi.org/10.1063/1.4927803}.

\bibitem{li_sciadv18_machine_learning_phase_transitions}
\bibinfo{author}{Li, L.} \emph{et~al.}
\newblock \bibinfo{journal}{\bibinfo{title}{Machine
  learning{\textendash}enabled identification of material phase transitions
  based on experimental data: Exploring collective dynamics in ferroelectric
  relaxors}}.
\newblock {\emph{\JournalTitle{Science Advances}}}
  \textbf{\bibinfo{volume}{4}}, \bibinfo{pages}{eaap8672},
  \doiprefix\url{10.1126/sciadv.aap8672} (\bibinfo{year}{2018}).

\bibitem{griffin_npjcompmat19_smart_machine_learning}
\bibinfo{author}{Griffin, L.~A.}, \bibinfo{author}{Gaponenko, I.},
  \bibinfo{author}{Zhang, S.} \& \bibinfo{author}{Bassiri-Gharb, N.}
\newblock \bibinfo{journal}{\bibinfo{title}{Smart machine learning or
  discovering meaningful physical and chemical contributions through
  dimensional stacking}}.
\newblock {\emph{\JournalTitle{npj Computational Materials}}}
  \textbf{\bibinfo{volume}{5}}, \doiprefix\url{10.1038/s41524-019-0222-z}
  (\bibinfo{year}{2019}).

\bibitem{gaponenko_rsi16_humidity_controller}
\bibinfo{author}{Gaponenko, I.}, \bibinfo{author}{Gamperle, L.},
  \bibinfo{author}{Herberg, K.}, \bibinfo{author}{Muller, S.~C.} \&
  \bibinfo{author}{Paruch, P.}
\newblock \bibinfo{journal}{\bibinfo{title}{Low-noise humidity controller for
  imaging water mediated processes in atomic force microscopy}}.
\newblock {\emph{\JournalTitle{Review of Scientific Instruments}}}
  \textbf{\bibinfo{volume}{87}}, \bibinfo{pages}{063709},
  \doiprefix\url{10.1063/1.4954285} (\bibinfo{year}{2016}).
\newblock \eprint{https://doi.org/10.1063/1.4954285}.

\bibitem{gaponenko_erx19_humidity_controller}
\bibinfo{author}{Gaponenko, I.}, \bibinfo{author}{Musy, L.},
  \bibinfo{author}{Muller, S.~C.} \& \bibinfo{author}{Paruch, P.}
\newblock \bibinfo{journal}{\bibinfo{title}{Open source standalone relative
  humidity controller for laboratory applications}}.
\newblock {\emph{\JournalTitle{Engineering Research Express}}}
  \textbf{\bibinfo{volume}{1}}, \bibinfo{pages}{025042},
  \doiprefix\url{10.1088/2631-8695/ab5771} (\bibinfo{year}{2019}).

\bibitem{Dumitrescu2018_dict_learning}
\bibinfo{author}{Dumitrescu, B.} \& \bibinfo{author}{Irofti, P.}
\newblock \emph{\bibinfo{title}{Dictionary Learning Algorithms and
  Applications}} (\bibinfo{publisher}{Springer International Publishing},
  \bibinfo{year}{2018}).

\bibitem{Tong2015}
\bibinfo{author}{Tong, S.} \emph{et~al.}
\newblock \bibinfo{journal}{\bibinfo{title}{Mechanical removal and rescreening
  of local screening charges at ferroelectric surfaces}}.
\newblock {\emph{\JournalTitle{Physical Review Applied}}}
  \textbf{\bibinfo{volume}{3}}, \doiprefix\url{10.1103/physrevapplied.3.014003}
  (\bibinfo{year}{2015}).

\bibitem{Rocoreanu2000_fitzhugh_nagumo}
\bibinfo{author}{Roc{\c{s}}oreanu, C.}, \bibinfo{author}{Georgescu, A.} \&
  \bibinfo{author}{Giurgi{\c{t}}eanu, N.}
\newblock \emph{\bibinfo{title}{The {FitzHugh}-Nagumo Model}}
  (\bibinfo{publisher}{Springer Netherlands}, \bibinfo{year}{2000}).

\bibitem{gariglio_apl_2007_pzt_growth}
\bibinfo{author}{Gariglio, S.}, \bibinfo{author}{Stucki, N.},
  \bibinfo{author}{Triscone, J.-M.} \& \bibinfo{author}{Triscone, G.}
\newblock \bibinfo{journal}{\bibinfo{title}{Strain relaxation and critical
  temperature in epitaxial ferroelectric pb(zr0.20ti0.80)o3 thin films}}.
\newblock {\emph{\JournalTitle{Applied Physics Letters}}}
  \textbf{\bibinfo{volume}{90}}, \bibinfo{pages}{202905},
  \doiprefix\url{10.1063/1.2740171} (\bibinfo{year}{2007}).
\newblock \eprint{https://doi.org/10.1063/1.2740171}.

\bibitem{rossant2018ipython}
\bibinfo{author}{Rossant, C.}
\newblock \emph{\bibinfo{title}{IPython Interactive Computing and Visualization
  Cookbook, Second Edition : Over 100 hands-on recipes to sharpen your skills
  in high-performance numerical computing and data science in the Jupyter
  Notebook}} (\bibinfo{publisher}{Packt Publishing},
  \bibinfo{address}{Birmingham}, \bibinfo{year}{2018}).

\bibitem{yareta}
\bibinfo{author}{Gaponenko, I.} \emph{et~al.}
\newblock \bibinfo{title}{{Local and correlated studies of humidity-mediated
  ferroelectric thin film surface charge dynamics}},
  \doiprefix\url{10.26037/yareta:wq6me5d75jb6vkz7rhva45ht6i}
  (\bibinfo{year}{2020}).

\end{thebibliography}

\section*{Data availability}
Data and analysis code is available at the long term storage Yareta project repository, hosted at the University of Geneva \cite{yareta}.

\section*{Acknowledgements}
The authors acknowledge Dr Sergei V. Kalinin of Oak Ridge National Laboratory, for helpful discussions about machine learning and the initial suggestion to explore reaction-diffusion modeling. This work was supported by Division II of the Swiss National Science Foundation under project 200021\_178782.

\section*{Author contributions}
The authors declare no competing financial interests.

I.G. performed the measurements and simulations on films grown by N.S. I.G. and L.M. analyzed the data. I.G. and P.P. wrote the manuscript. All authors contributed to the scientific discussion and manuscript revisions. 

\section*{Supplementary Materials}
\subsection*{Complete humidity-sample-time dataset}
The complete humidity-sample-time dataset is shown in Fig. \ref{fig:smdataset}, for the two samples (up and down polarised) and ten humidities.
\renewcommand{\thefigure}{S1}
\begin{figure}[ht]
\centering
\includegraphics[width=0.7\linewidth]{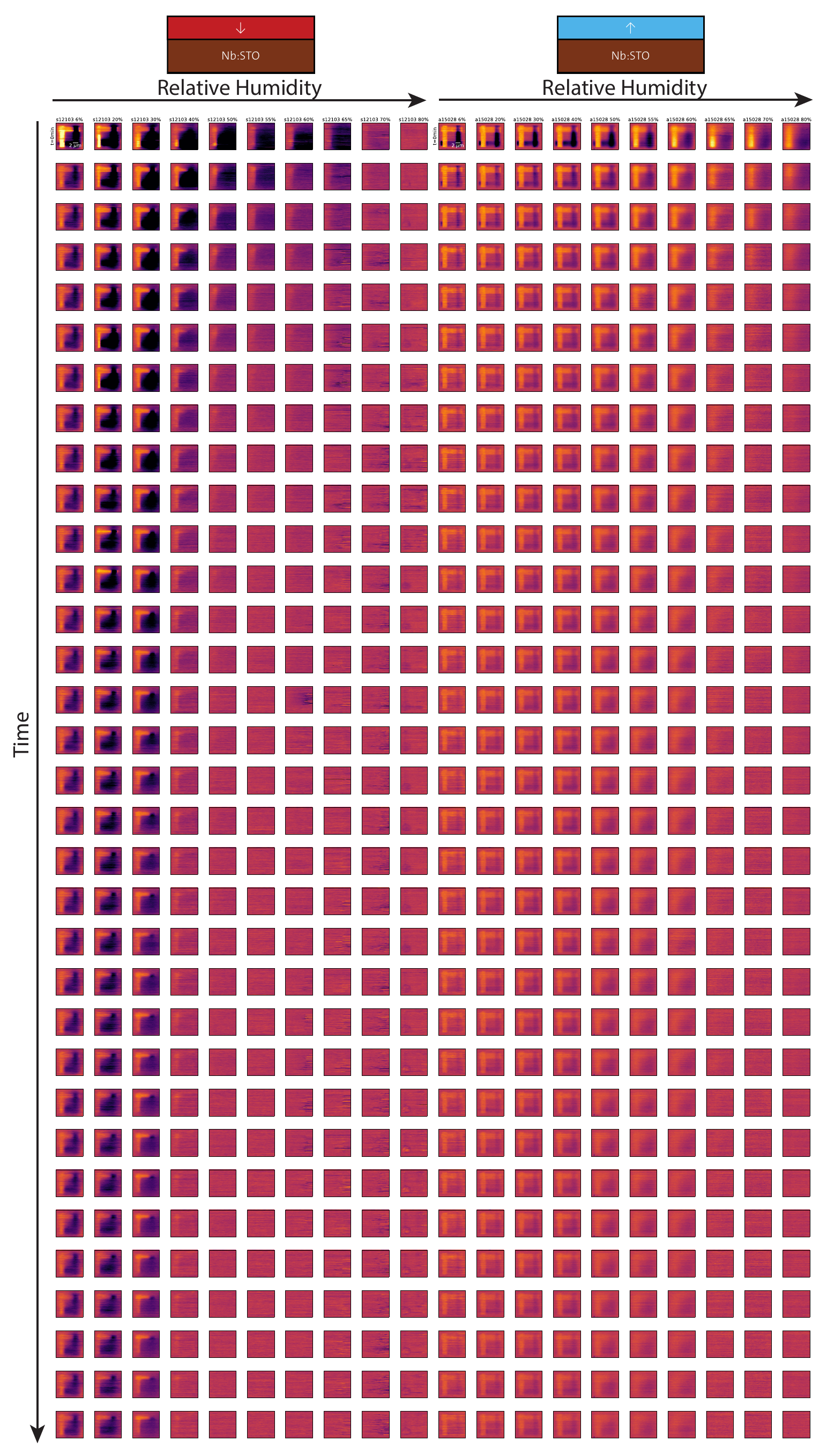}
\caption{Complete KPFM dataset as a function of sample, relative humidity and time.}
\label{fig:smdataset}
\end{figure}
\subsection*{Spectral decomposition principle}
The principle behind spectral decomposition methods such as DictionaryLearning is shown in Fig. \ref{fig:smprinciple}. The complete dataset is split into a set of one-dimensional components with corresponding two-dimensional weight maps. Dimensional stacking was used to integrate the relative humidity and initial sample polarisation dependence into the maps, thus correlating the data sets and enabling quantitative comparison of time-dependent decays.
\renewcommand{\thefigure}{S2}
\begin{figure}[ht]
\centering
\includegraphics[width=\linewidth]{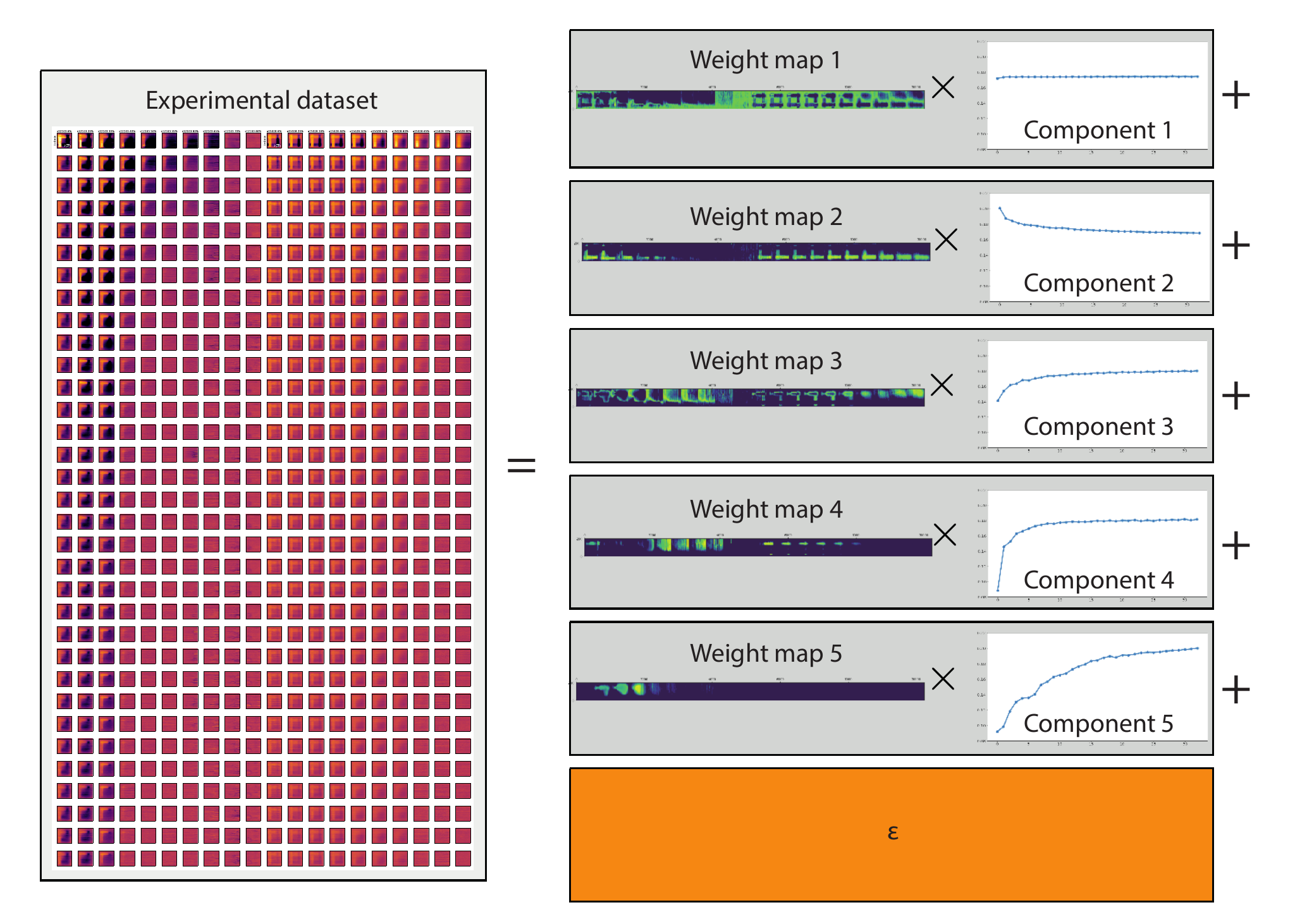}
\caption{Principle behind machine learning based spectral decomposition.}
\label{fig:smprinciple}
\end{figure}
\subsection*{Dependence on the number of components}
A systematic investigation of the influence of the number of components on the DictionaryLearning analysis is shown in Fig. \ref{fig:smcomponents}. As the number of components increases, so does the number of physical behaviours. At first, when $n=2$, only a single physical behaviour is visible, alongside a Gaussian noise component with a null weight map. When the number of allowed components is increased up to $n=6$, the negative voltage process splits from the main component. Increasing the number of components to $n=7$, the concentric structure for the negative voltage region starts to form with the split of the latter into two distinct map/component pairs. At $n=8$, a background component splits from the positive voltage component, and finally at $n=10$, a third concentric negative component makes its appearance.
\renewcommand{\thefigure}{S3}
\begin{sidewaysfigure}[ht]
\centering
\includegraphics[width=\linewidth]{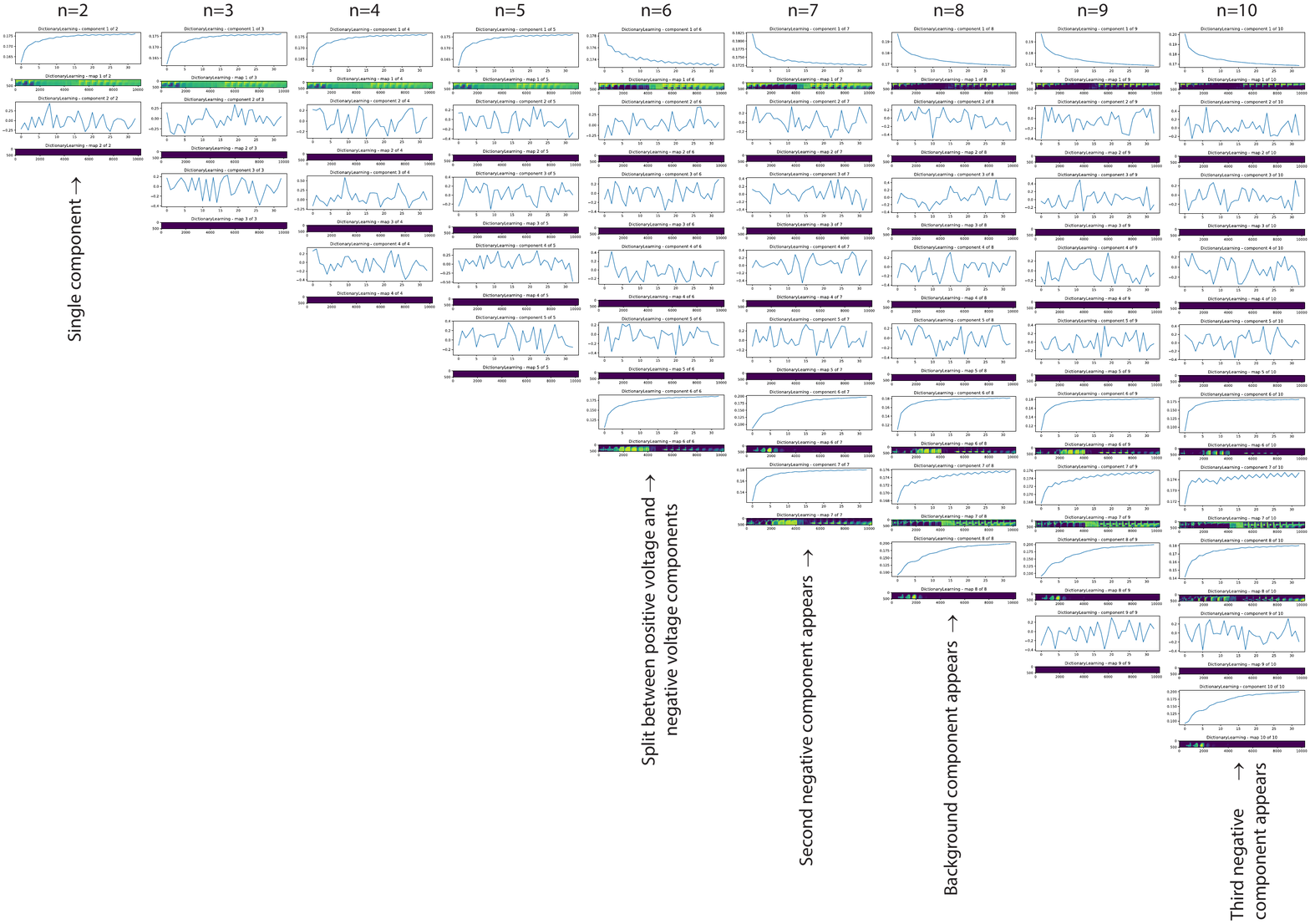}
\caption{Dictionary Learning based spectral decomposition of the correlated KPFM dataset as a function of the number of components. Increasing number of components refines the separation into further sub-processes.}
\label{fig:smcomponents}
\end{sidewaysfigure}
\subsection*{Reaction-diffusion modeling}
The reaction-diffusion modeling, here based on a Fitzhugh-Nagumo excitation-relaxation process, provides insight into the contribution of the various physical processes to the machine learning based analysis. Such a sandbox approach enables a unique way to tailor the interactions present within the two species. In particular, the reaction-diffusion modeling highlights the importance of sparse approaches such as DictionaryLearning when dealing within physical materials presenting very different co-located behaviors.

\subsubsection*{Initial conditions choice}
Initial species distributions were chosen to mimic the positive and negative charges as deposited by biased-tip atomic force microscopy writing. The positive charge species $u$ shown in Fig. \ref{fig:smrdinit}(a) was initialized with only a reaction interaction with the negative charge species $v$, and no diffusion component. The negative charge species $v$, shown in Fig. \ref{fig:smrdinit}(b) is susceptible to both reaction and diffusion processes. The total distribution $u-v$ is shown in Fig. \ref{fig:smrdinit}(c), and qualitatively resembles the Kelvin probe force microscopy data in Fig. \ref{fig:kpfm}.
\renewcommand{\thefigure}{S4}
\begin{figure}[ht]
\centering
\includegraphics[width=0.75\linewidth]{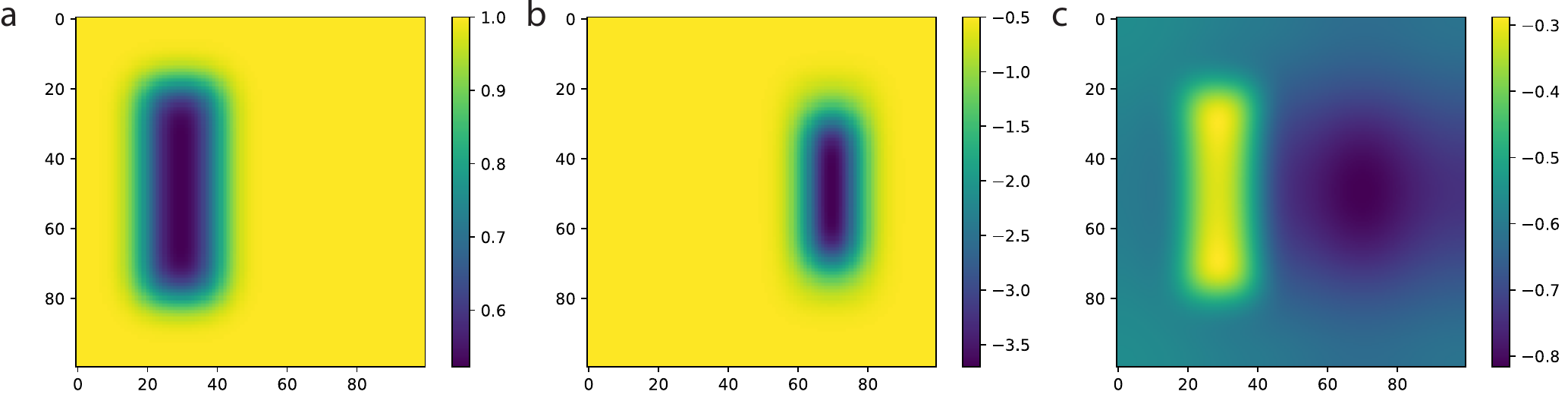}
\caption{Initial distributions of the two reacting/diffusing species (a) $u$ and (b) $v$. The resulting potential is computed in (c) as $u-v$ and processed further with machine learning techniques.}
\label{fig:smrdinit}
\end{figure}
\subsubsection*{Principal component analysis}
Whilst principal component analysis (PCA) does not provide a physically interpretable spectral decomposition, it can be used to estimate the complexity of the decomposition. The PCA weight maps for the first 50 components are shown in Fig. \ref{fig:smrdpca}(a), illustrating the concentric-like pattern generated by the diffusive nature of the negative charge species distribution $v$. Indeed, the positive charge species distribution $u$ is only present in the first four maps of the decomposition. The corresponding PCA components shown in Fig. \ref{fig:smrdpca}(b) additionally shows a periodic behaviour with increasing frequency for the first 26 components, as evidenced by their Fourier transform in Fig. \ref{fig:smrdpca}(c).
\renewcommand{\thefigure}{S5}
\begin{figure}[ht]
\centering
\includegraphics[width=\linewidth]{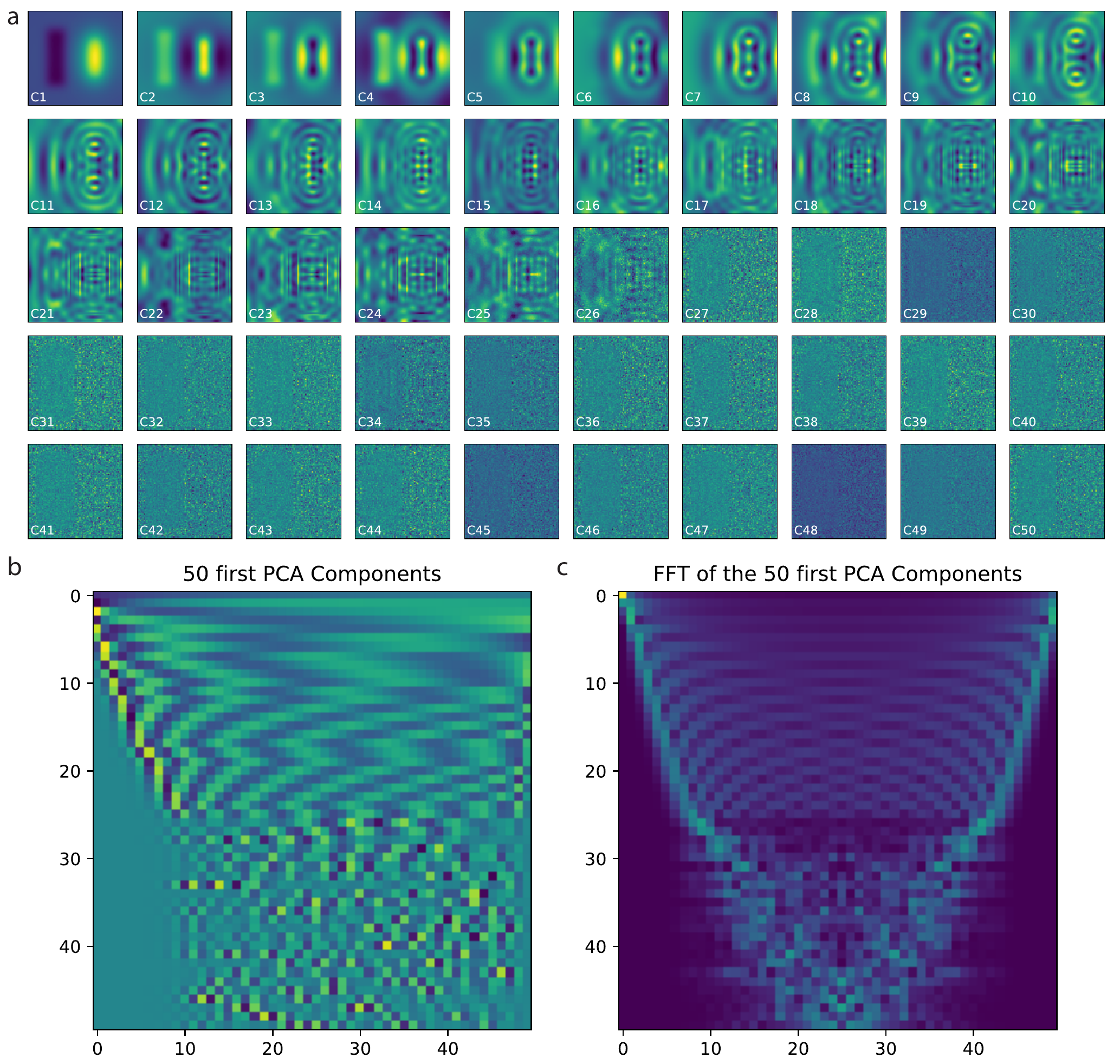}
\caption{Principal component analysis of the reaction-diffusion dataset up to 50 components, with (a) the maps and (b) their corresponding components. (c) The Fourier transform in the components shows an increasing oscillation frequency for the increasing order components.}
\label{fig:smrdpca}
\end{figure}
\subsubsection*{Non-negative matrix factorization}
A non-negative matrix factorization (NMF) approach was tried concurrent to DictionaryLearning, with results for $n=4$ and $n=10$ components shown in Fig. \ref{fig:smrdnmf}(b,c) respectively. Although not able to separate the physical processes as cleanly as DictionaryLearning, NMF also highlights the concentric decomposition of the negative voltage species with an increasing number of components.
\renewcommand{\thefigure}{S6}
\begin{figure}[ht]
\centering
\includegraphics[width=\linewidth]{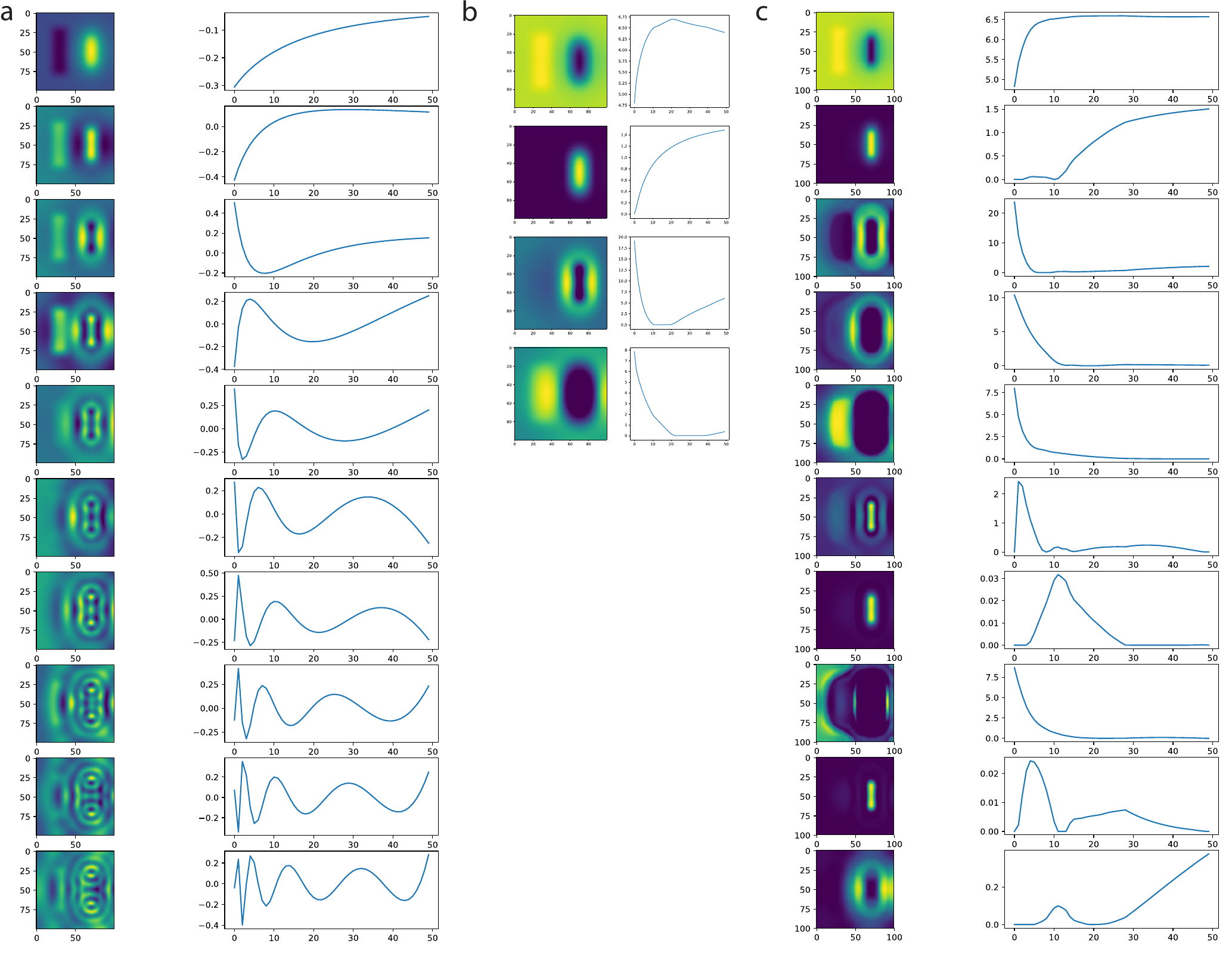}
\caption{Comparison between PCA (a) and NMF (b) $n=4$ and (c) $n=10$ methods. In all cases - just like in the DictionaryLearning analysis - the negative voltage species generate concentric components due to their diffusive nature.}
\label{fig:smrdnmf}
\end{figure}
\end{document}